\documentclass[twocolumn,prl,showpacs,floatfix]{revtex4}
\usepackage{dcolumn}
\usepackage{graphicx}
\usepackage{epsf}
\usepackage{amsmath}
\usepackage{bm}

\newcommand{\addrFreiburg}{Fakult\"{a}t f\"{u}r Mathematik und Physik der 
Albert--Ludwigs--Universit\"{a}t, Theoretische Quantendynamik,\\
Hermann--Herder--Stra\ss{}e 3, D--79104 Freiburg, Germany}

\begin{document}

\title{Lamb Shift of Laser-Dressed Atomic States}

\author{Ulrich D.~\surname{Jentschura}}
\email{ulj@physik.uni-freiburg.de}

\author{J\"{o}rg \surname{Evers}}
\email{evers@physik.uni-freiburg.de}

\author{Martin \surname{Haas}}
\email{haas@physik.uni-freiburg.de}

\author{Christoph H. \surname{Keitel}}
\email{keitel@uni-freiburg.de}

\affiliation{\addrFreiburg}


\begin{abstract}
We discuss radiative corrections to an atomic two-level system
subject to an intense driving laser field. It is shown that 
the Lamb shift of the laser-dressed states, which are the
natural state basis of the combined atom-laser system, cannot
be explained in terms of the Lamb shift received by the atomic
bare states which is usually observed in spectroscopic experiments. 
In the final part, we propose an experimental scheme to measure these
corrections based on the incoherent resonance fluorescence spectrum
of the driven atom.
\end{abstract}


\maketitle
The interaction of coherent light with matter
is of cardinal interest both from a theoretical
point of view as well as for applications. 
Thus it is not surprising that different approaches
to this problem have been proposed and successfully applied.
At the most fundamental level, 
quantum electrodynamics (QED) is one of the most accurate theories
known so far~\cite{ItZu1980,MoTa2000,SaYe1990}. 
The bound-state 
self-energy as predicted by QED is the dominant radiative correction
in hydrogenlike systems and gives $98\,\%$ 
of the ground-state Lamb shift in atomic hydrogen~\cite{JeMoSo1999}.
QED radiative corrections
are usually evaluated with the adiabatic $S$--matrix 
formalism~\cite{GMLo1951,MoPlSo1998}.
A complementary approach to the matter-light interaction 
based on the same foundations
is quantum optics~\cite{CTDRGr1989,CTDRGr1992,ScZu1997},
which is 
especially suited for the description of time-dependent 
processes for which the adiabatic $S$--matrix is of limited use.
Within the quantum optical formalism,
the atom-laser interaction may intuitively be understood
with the help of the so-called dressed states which are
defined as the eigenstates of the interacting system of light
and matter~\cite{CT1975}. 
A textbook application
is the incoherent, inelastic 
resonance fluorescence spectrum of an atomic system subject to
a driving laser field~(see~\cite[Ch.~10]{ScZu1997}).
It is well known that the peaks of the incoherent spectrum 
may easily be interpreted with the help
of dressed states. The incoherent fluorescence 
has received considerable attention in 
the past, both theoretically and experimentally, 
as it may be modified by external influences to a great extent
(see~\cite[Ch.~10]{ScZu1997}) and references therein).

In this Letter we take advantage of ideas from quantum electrodynamics
and from quantum optics to analyze radiative corrections received by 
laser-dressed atomic states. 
This constitutes the rather fundamental
field-theoretic problem of the electron-to-vacuum interaction 
with the electron being bound to an atomic nucleus
and being driven simultaneously by an additional strong external
laser field.
From the viewpoint of QED, our setup corresponds to the 
strong-coupling regime of the atom to the laser field.
This should be distinguished from the strong binding (Coulomb) field 
limit of the Lamb shift usually found in high--$Z$ ions~\cite{MoPlSo1998}, 
from the radiative shifts of Volkov states~\cite{EbRe1966}, and from 
radiative corrections in modified vacuum structures such as in photonic 
crystals~\cite{ZhYa2000}. 
The dominant interaction in the system is the coupling of the atom 
to the driving laser field which gives rise to the atomic laser-dressed 
states. This interaction is taken into account to all orders in the 
atom-laser coupling within the rotating-wave approximation (RWA).
Starting from the natural dressed state basis of the system,
we perturbatively calculate the self-energy corrections
(as we focus on the self-energy, we will use the terms 
``Lamb shift'' and ``self-energy'' interchangeably).
Thus we first incorporate the strong interaction 
with the laser and treat the second-order shift due to the 
vacuum field in a second step of the calculation.
This way we find that the self-energy shift
of the laser-dressed states clearly deviates in a nontrivial manner
from the usual $S$-matrix results for atomic bare states.
We further point out situations where the modified radiative 
corrections are also of practical relevance.

The system under consideration is a monochromatic laser field
which couples near-resonantly to an electric-dipole allowed 
transition $|e\rangle \leftrightarrow |g\rangle$ of a single atom.
In a typical quantum optical treatment in two-level-, dipole- and 
RWA (see~\cite[Ch.~10]{ScZu1997}), the system Hamiltonian may 
be approximated as ($\hbar = c = \epsilon_0 = 1$)
\begin{eqnarray}
\label{HRWA}
{\cal H}_{\rm RWA} &=& 
\omega_g \, | g \rangle \, \langle g | + 
\omega_e \, | e \rangle \, \langle e | +  
\omega_{\rm L} a^\dagger_{\rm L} \, a_{\rm L}  \nonumber\\
&& + g_{\rm L} \,
\left( a_{\rm L}^\dagger \, | g \rangle \, \langle e | 
 + a_{\rm L} \, | e \rangle \, \langle g | \right) \,.
\end{eqnarray}
The $\omega_i$ ($i=e,g$) are the energies of the respective atomic states,
$\omega_{\rm L}$ is the frequency of the laser field, $a_{\rm L}$ 
($a^\dagger_{\rm L}$)
are photon annihilation (creation) operators for the laser field mode,
and $g_{\rm L}$ is a coupling constant.

The driving of the external laser field gives rise to a
resonance fluorescence spectrum which consists of an elastic scattering part centered at the frequency  of the driving laser field and an incoherent part, which for $\Omega_n \gg \Gamma$ (secular limit) splits up into three distinct 
peaks. Here, $\Omega_n = 2\,g_{\rm L}\,\sqrt{n+1}$ is the Rabi 
frequency of the driven transition which
depends on the number of photons $n$ in the laser field mode and
$\Gamma$ is the decay rate of the transition. 
The main peak of this Mollow spectrum
is again centered at the driving laser field frequency,
while the two other peaks are shifted by the generalized
Rabi frequency $\pm \Omega_{\rm R}^{(n)} = \pm \sqrt{\Omega_n^2 + \Delta^2}$ 
to higher and lower frequencies, respectively,
with $\Delta = \omega_L - \omega_{\rm R}$ as the 
detuning of the driving laser field ($\omega_{\rm R} = \omega_e - \omega_g$).

The dressed states~\cite{CT1975} are the eigenstates of the combined system of 
the atomic two-level system and the driving laser field in RWA and may
be written as 
\begin{subequations}
\label{dressedstates}
\begin{eqnarray}
\label{dressedstateA}
|(+,n)\rangle &=& \cos\theta_n \, | e, n\rangle +
\sin\theta_n \, | g, n+1\rangle\,,
\\[2ex]
\label{dressedstateB}
|(-,n)\rangle &=& -\sin\theta_n \, | e, n\rangle +
\cos\theta_n \, | g, n+1\rangle\,.
\end{eqnarray}
\end{subequations}
Here, 
$|i, n\rangle$ ($i\in\{e,g\}$) denotes the state where the atom is in 
the bare level $i$ 
with $n$ photons in the driving laser field mode, and
$\theta_n$ is the mixing angle defined by 
$\tan (2 \theta_n) = - \Omega_n / \Delta$.
The energies of these dressed states in RWA are given by
\begin{equation}
\label{dressedenergies}
E_{\pm,n} = \left( n + \frac12 \right) \, \omega_{\rm L} +
\frac12\, \omega_{\rm R} \pm \frac12\, \Omega^{(n)}_{\rm R}\,.
\end{equation}
The Mollow spectrum may then be understood as originating
from transitions $|(\pm,n)\rangle \rightarrow |(\pm,n-1)\rangle$
among the dressed states.
As the driving laser field discussed here is sufficiently intense,
we replace $\Omega_n$, $\Omega^{(n)}_{\rm R}$and $\theta_n$ by their corresponding semiclassical entities $\Omega$, $\Omega_{\rm R}$
and $\theta$ in the following discussion.

The Hamiltonian ${\cal H}_{\rm R}$ describes the interaction of the atom with
all modes but the laser field mode, and ${\cal H}_{\rm F}$
describes the electromagnetic field,
\begin{equation}
\label{HRHF}
{\cal H}_{\rm R} = -q \, \bm{r} \cdot \bm{E}_{\rm R}\,, 
\quad
{\cal H}_{\rm F} =
\sum_{\bm{k}\lambda}
\omega_{\bm{k}}
a^+_{\bm{k}\lambda} \, a_{\bm{k}\lambda}\,.
\end{equation}
Here, $q$ is the physical charge of the electron ($q^2 = 4 \pi \alpha$ where
$\alpha$ is the fine-structure constant), and $\bm{r}$ is the 
position operator.
The electric field operator of the non-laser modes is given by
\begin{equation}
\bm{E}_{\rm R} =
\sum_{\bm{k}\lambda \neq L}
\sqrt{\frac{\omega_{\bm k}}{2\,V}} \,
\epsilon_\lambda(\bm k) \,
\left[
a_{\bm{k}\lambda} +
a^\dagger_{\bm{k}\lambda} \,
\right]  \,.
\end{equation}
$V$ is the quantization volume, $\epsilon_\lambda(\bm k)$ is a polarization
vector, and $\omega_{\bm k}$, 
$a_{\bm{k}\lambda}$ and $a^\dagger_{\bm{k}\lambda}$ are the
frequency, the annihilation and the creation operator of the vacuum mode
with wave vector $\bm k$ and polarization $\lambda$, respectively.

The second-order radiative self-energy shift 
arises from two terms. First, we have
the non-laser-field radiation modes and resonant
intermediate atomic states (treated as dressed states within the RWA)
\begin{equation}
\label{L1}
\Delta L^{(1)}_{\pm,n} = \left< (\pm, n) \left| {\cal H}_{\rm R} 
\frac{1}{E_{\pm,n} - {\cal H}_{\rm res}} 
{\cal H}_{\rm R} \right| (\pm,n) \right>\,,
\end{equation}
with ${\cal H}_{\rm res} = {\cal H}_{\rm RWA} + 
{\cal H}_{\rm F}$. Second, we consider
off-resonant intermediate states,
\begin{equation}
\label{L2}
\Delta L^{(2)}_{\pm,n} =
\left< (\pm, n) \left| {\cal H}_{\rm R} \,
\frac{1}{E_{\pm,n} - {\cal H}_{\rm off}} \,
{\cal H}_{\rm R} \right| (\pm, n) \right>\,,
\end{equation}
where ${\cal H}_{\rm off}$ is given by ${\cal H}_{\rm res}$ 
under the replacement ${\cal H}_{\rm RWA} \to
\sum_{j \neq g,e} \omega_j \, | j \rangle \, \langle j |$,
excluding the resonant states
$|e\rangle$, $|g\rangle$.

It is natural to assume that in the limit of vanishing laser
intensity $\Omega_{\rm R} \to 0$ and vanishing detuning 
$\Delta \to 0$, the Lamb shift of the dressed states should be equal to the 
radiative shift we would expect from the usual bare-state treatment 
of the Lamb shift~\cite{SaYe1990,MoPlSo1998}. Indeed, 
neglecting the detuning and the Rabi frequency,
the sum of the terms (\ref{L1})$+$(\ref{L2}) leads to
the following approximative (app) result
\begin{eqnarray}
\label{LplusApproxAgain}
\lefteqn{\Delta L^{(\rm app)}_{+,n} =
\frac{4\alpha}{3 m^2} \, (Z\alpha) \,
\ln[(Z\alpha)^{-2}] }
\nonumber\\
& & \times \left\{
\cos^2\theta \, \langle e | \delta^{(3)}(\bm{r}) | e \rangle +
\sin^2\theta \, \langle g | \delta^{(3)}(\bm{r}) | g \rangle
\right\}\,,
\end{eqnarray}
and the shift $\Delta L^{(\rm app)}_{-,n}$
of $\omega_-$ is obtained by replacing
$\sin\theta \leftrightarrow \cos\theta$ in the above formula
($Z$ is the nuclear charge number, and $m$ is the electron mass).
This result may be rewritten as
\begin{equation}
\label{LApprox}
\Delta L^{(\rm app)}_{\pm,n} = 
\left< (\pm,n) | \Delta V_{\rm Lamb}(r) | (\pm,n) \right>
\end{equation}
with $\Delta V_{\rm Lamb}(\bm{r}) =
4 \alpha \, (Z\alpha) \, \ln[(Z\alpha)^{-2}] \,
\delta^{(3)}(\bm{r}) /(3 \, m^2)\,.$
The expectation value vanishes 
of this approximative ``effective Lamb shift'' 
potential~\cite{Ka1996,JeNa2002} vanishes for all states with angular
momentum $l \geq 1$~\cite{RemarkLambShift}. 

We now investigate the corrections to the shift
of the high- and low-frequency Mollow sidebands $\Delta \omega_\pm$
due to Eq.~(\ref{LApprox}) with respect to the lowest-order results
\begin{equation}
\omega_+ = E_{+,n} - E_{-,n-1}\,, \qquad
\omega_- = E_{-,n} - E_{+,n-1}\,.
\end{equation}
We obtain
\begin{equation}
\label{Deltaomegaplus}
\Delta \omega_+ = 
\Delta L^{(\rm app)}_{+,n} - \Delta L^{(\rm app)}_{-,n-1} 
= -\frac{\Delta}{\sqrt{\Omega^2 + \Delta^2}} \,
L_{\rm bare}
\end{equation}
where
$
L_{\rm bare} = \langle e | \Delta V_{\rm Lamb} | e \rangle -
\langle g | \Delta V_{\rm Lamb} | g \rangle
$
is the effective Lamb shift acquired by the bare states. Also,
we have $\Delta \omega_- = - \Delta \omega_+$.

When we keep the terms linear in $\Omega_{\rm R}$ and 
$\Delta$ in evaluating the matrix elements
in Eqs.~(\ref{L1})$+$(\ref{L2}), we obtain the following
corrections $\Delta C_{\pm,n}$
to the leading-order shift of the dressed states
$|(\pm,n)\rangle$ given in Eq.~(\ref{LApprox})
[for a detailed derivation we refer the reader to~\cite{JeKe2003}]:
\begin{subequations}
\begin{eqnarray}
\lefteqn{\Delta C_{+,n} =
-\frac{\alpha}{\pi} \,
\ln[(Z\alpha)^{-2}] \,
\frac{1}{m^2} }  \nonumber \\
&&\: \times \bigg [ \cos^2\theta \left < \bm{p}^2 \right>_e 
(\Omega_{\rm R} + \Delta) 
+ \sin^2\theta \left< \bm{p}^2 \right>_g 
(\Omega_{\rm R} - \Delta)  \nonumber\\
&&\:  + \left| \left< \bm{p} \right>_{eg} \right|^2 
(\Delta \cos (2\theta)  + \Omega_{\rm R} \cos^2 (2\theta) ) \bigg ] \,,
\label{deltacplus}
\\[2ex]
\lefteqn{\Delta C_{-,n} =
\frac{\alpha}{\pi} \,
\ln[(Z\alpha)^{-2}] \,
\frac{1}{m^2} }  \nonumber \\
&&\: \times \bigg [ \cos^2\theta \left < \bm{p}^2 \right>_g 
(\Omega_{\rm R} + \Delta) 
+ \sin^2\theta \left< \bm{p}^2 \right>_e 
(\Omega_{\rm R} - \Delta)  \nonumber\\
&&\:  + \left| \left< \bm{p} \right>_{eg} \right|^2 
(\Delta \cos (2\theta)  + \Omega_{\rm R} \cos^2 (2\theta) ) \bigg ] \,.
\label{deltacminus}
\end{eqnarray}
\end{subequations}
Here $\left< \bm{p} \right>_{ij} = \left< i \left| \bm{p} \right| j \right>$
is the dipole matrix element, 
and $\left< \bm{p}^2 \right>_j = \left< j \left| \bm{p}^2 \right| j \right>$
is the expectation value of the square of the atomic momentum where 
$|i\rangle$ and $|j\rangle$ denote atomic bare states.

The additional shift to the high- and low-frequency Mollow sidebands
$\omega_\pm$ due to Eqs.~(\ref{deltacplus}) and (\ref{deltacminus}), 
which we denote by $\delta \omega_\pm$ in contrast to 
$\Delta \omega_\pm$, may be simplified to~\cite{JeKe2003}
\begin{equation}
\label{deltaomegaplusminus}
\delta \omega_\pm =
\mp {\cal C}\, \frac{\Omega^2}{\sqrt{\Omega^2 + \Delta^2}}\,,\,\,\,
{\cal C} = \frac{\alpha}{\pi} \, \ell \,
\frac{\left< \bm{p}^2 \right>_g + \left< \bm{p}^2 \right>_e}{m^2}\,.
\end{equation}
Here, $\ell = \ln[(Z\alpha)^{-2}]$, and ${\cal C}$ is dimensionless.

The Mollow sidebands are thus Lamb shifted in total 
according to Eqs.~(\ref{Deltaomegaplus})
and (\ref{deltaomegaplusminus}) by
\begin{eqnarray}
\label{shift}
\lefteqn{\omega_+ \to \omega_+ + \Delta \omega_+ + \delta \omega_+ }
\\
&=& \omega_{\rm L} + \sqrt{\Omega^2 + \Delta^2} 
- \frac{\Delta}{\sqrt{\Omega^2 + \Delta^2}} \, L_{\rm bare}
- {\cal C}\, \frac{\Omega^2}{\sqrt{\Omega^2 + \Delta^2}}
\nonumber\\
&=& \omega_{\rm L} + \sqrt{\Omega^2 \, (1 - {\cal C})^2 +
(\Delta - L_{\rm bare})^2} + {\cal O}(\Omega^2, \Delta^2) \,.
\nonumber
\end{eqnarray}
It is highly suggestive to interpret the approximative 
Lamb shift $\Delta \omega_\pm$ as generating,
under the square root, a specific term that effectively
shifts the detuning $\Delta$ by an amount that corresponds to the 
Lamb shift of the bare transition,
$\left( \omega_{\rm R} \to \omega_{\rm R} + L_{\rm bare}\right)  
\Leftrightarrow 
\left( \Delta \to \Delta - L_{\rm bare}\right)$.
This correction is therefore not a genuinely new effect; it can
be obtained without invoking the dressed-state formalism and 
can be included {\em a posteriori} in evaluating the 
theoretical value of the detuning with the ``bare'',
or in other words ``usual'', Lamb shift being taken into account.

In contrast, the shift mediated by the ${\cal C}$-term
in Eq.~(\ref{shift}), which is effectively a radiative modification
of the Rabi frequency, cannot be obtained unless we use the 
dressed-state formalism. We are therefore led
to define the {\em fully dressed Lamb shift} of 
the two Mollow sidebands as
\begin{eqnarray}
\label{fullydressed}
\Delta^{(\rm full)} {\cal L}_\pm &=& \pm
\left( \sqrt{\Omega^2 \, (1 - {\cal C})^2 +
(\Delta - L_{\rm bare})^2}\right.
\nonumber\\[2ex]
& & \left. - \sqrt{\Omega^2 + \Delta^2}\right)
\approx \Delta\omega_\pm + \delta\omega_\pm\,.
\end{eqnarray}

We now turn to the experimental 
verification of the radiative corrections to the Mollow
spectrum. A precision
measurement of the Mollow spectrum is required.
The atomic system under
study should be described to very good accuracy by the two-level
approximation. Otherwise, considerable further complications
due to a multi-level formalism would arise.
A further prerequisite is a frequency- and intensity-stabilized
continuous-wave (cw) laser tuned to the atomic resonance to
allow the system to evolve into the steady-state.

We recall the explicit familiar three-peak Mollow spectrum
which describes the frequency-dependent 
intensity spectrum of the incoherent fluorescence
(secular limit),
\begin{eqnarray}
\label{MollowSpectrumSecular}
\lefteqn{{\cal S}_{\rm inc}(\omega) \approx
\frac{\Gamma}{\pi}\, \left[
\frac{\Gamma_0 \, A^{\rm inc}_0}
  {(\omega - \omega_{\rm L})^2 + \Gamma^2_0}
\right.}
\\[2ex]
& & \left. + \frac{\Gamma_+ \, A_+}
  {(\omega - \omega_{\rm L} - \Omega_{\rm R})^2 + \Gamma_+^2}
+ \frac{\Gamma_- \, A_-}
  {(\omega - \omega_{\rm L} + \Omega_{\rm R})^2 + \Gamma_-^2}
\right]\,.
\nonumber
\end{eqnarray}
Corrections beyond the secular approximation may be expressed
as a series in $\Gamma/\Omega_{\rm R}$~\cite{JeKe2003}.
Modifications of the Mollow spectrum due to modified decay rates 
such as in a squeezed vacuum~\cite{CaLaWa1987}, via quantum 
interferences \cite{EvKe2002}  
as well as via modifications in strong driving fields with a 
Rabi frequency nonnegligible to that of the transition 
frequency~\cite{BrKe2000} have been discussed in the literature.

The generalized Rabi frequency in this formula becomes
\begin{equation}
\label{OmegaR}
\Omega_{\rm R} = \sqrt{\Omega^2 + \Delta^2}
\to \sqrt{\Omega^2 \, (1 - {\cal C})^2 + (\Delta - L_{\rm bare})^2}\,,
\end{equation}
in order to take care of both the bare Lamb shift and 
the radiative shift of the Rabi frequency, and the parameters in 
(\ref{MollowSpectrumSecular}) read:
\begin{eqnarray}
\label{Ainc0}
A^{\rm inc}_0 &=& \frac{\Omega^6}
  {4 \, \Omega_{\rm R}^2 \, (\Omega_{\rm R}^2 + \Delta^2)^2}\,,\,\,
A_\pm = \frac{\Omega^4}
  {8 \, \Omega_{\rm R}^2 \, (\Omega_{\rm R}^2 + \Delta^2)}\,,
\nonumber\\[2ex]
\label{Gamma0}
\Gamma_0 &=& \Gamma \, \frac{\Omega^2 + 2\,\Delta^2}{2 \, \Omega_{\rm R}^2}
\,,\,\,
\Gamma_\pm =
\Gamma \, \frac{3 \, \Omega^2 + 2\,\Delta^2}{4 \, \Omega_{\rm R}^2}\,.
\end{eqnarray}
Here, $\Gamma$ is the decay width of the upper atomic level 
$|e\rangle$ which also determines the width 
of the Mollow sidebands. Let us consider a situation
with vanishing detuning $\Delta$ (this implies
$\Omega = \Omega_{\rm R}$). Further, we define the 
ratio $h = \Omega/\Gamma$. The width of the 
Mollow sidebands $\Gamma_\pm$ is of the order of $\Gamma$ 
according to (\ref{Gamma0}). The radiative Rabi-frequency 
correction to the Mollow sidebands $\delta \omega_\pm$ 
is of the order of ${\cal C}\,\Omega$ 
(see Eq.~(\ref{deltaomegaplusminus})). We compare
$\delta \omega_\pm$ with the width of the Mollow sideband peak;
this leads to the following order-of-magnitude estimate (``$\sim$'') for the 
``shift-to-width'' ratio $r_1$, 
\begin{equation}
\label{r1}
r_1 = \frac{\delta\omega_\pm}{\Gamma} \sim h \, {\cal C}\,.
\end{equation}
The Bloch--Siegert shift $\delta_{\rm BS}\omega_\pm$ (see~\cite{BlSi1940})
of the dressed states is a second-order effect in the 
atom-laser interaction which at $\Delta=0$ shifts the dressed states by
a frequency of the order of $\Omega^3/\omega_{\rm L}^2$~\cite{BrKe2000}
(a formula valid for arbitrary detuning
is contained in~\cite{JeKe2003}). 
It is perhaps worth noting that according to~\cite{CTDRGr1992},
the Bloch--Siegert correction could therefore be interpreted as a
stimulated radiative correction. The ratio 
$r_2$ of the radiative shift $\delta\omega_\pm$ of the generalized 
Rabi-frequency to the Bloch--Siegert shift is
\begin{eqnarray}
\label{r2}
r_2 &=& \frac{\delta\omega_\pm}{\delta_{\rm BS} \omega_\pm} \sim
\frac{{\cal C}\,\Omega}{\Omega^3/\omega_{\rm L}^2} = 
\frac{\omega_{\rm L}^2\,{\cal C}}{\Omega^2} =
\frac{\ln[(Z\alpha)^{-2}]}{\alpha (Z \alpha)^2} \, h^{-2}\,.
\nonumber
\end{eqnarray}
We perform order-of-magnitude estimates
based on the $Z\alpha$-expansion~\cite{BeSa1957}.
The laser frequency (= atomic transition frequency) is 
$\omega_{\rm L} \sim (Z\alpha)^2\,m$,
the decay width is $\Gamma \sim \alpha\,(Z\alpha)^4\,m$, and 
${\cal C} \sim \alpha\,(Z\alpha)^2\,\ln[(Z\alpha)^{-2}]$ 
is defined in Eq.~(\ref{deltaomegaplusminus}).
With $h \approx 1000$ and ${\cal C} \sim 
\alpha\,(Z\alpha)^2\,\ln[(Z\alpha)^{-2}] \sim 10^{-6}$
(at $Z=1$), we obtain $r_1 \sim 10^{-3}$ and $r_2 \sim 10$.
A resolution of the peak of a Lorentzian to one part in 
$10^3$ of its width is feasible as well as the theoretical description
of the Bloch--Siegert shifts to the required accuracy~\cite{JeKe2003}.

\begin{figure}[t]
\includegraphics[height=2.8cm]{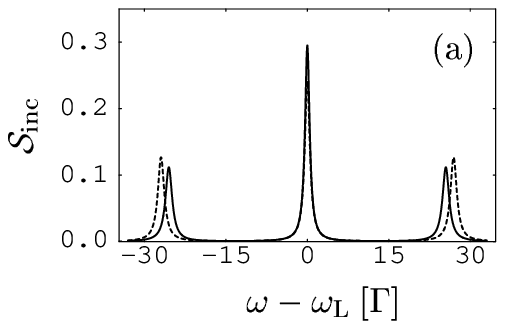}
\hspace*{0.1cm}
\includegraphics[height=2.8cm]{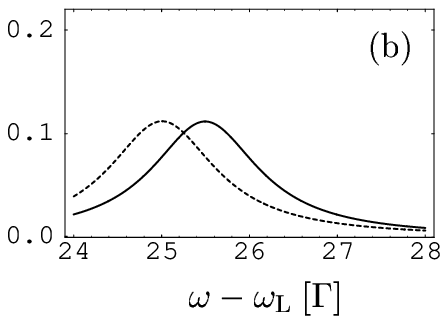}
\caption{\label{fig1}
In both figures (a) and (b), the solid line 
shows the Mollow spectrum (\ref{MollowSpectrumSecular}) 
corrected by the bare Lamb shift
according to Eq.~(\ref{LApprox}).
In Fig.~(a), the dashed line shows in addition the
Mollow spectrum (\ref{MollowSpectrumSecular})
without any Lamb shift.
In Fig.~(b), the dashed line shows the Mollow spectrum 
corrected by the fully dressed Lamb shift (see Eqs.~(\ref{shift}) 
and (\ref{fullydressed})).
For illustrational purposes, the parameters $\Gamma=1, \Omega=25, 
\Delta=10, \mathcal{C}=0.02$ and $L_{\rm bare}=5$ have been chosen.}
\end{figure}

The recently developed continuous-wave (cw) Lyman-alpha 
source~\cite{EiWaHa2001} was originally designed
to cool antihydrogen. We propose a measurement
on the hydrogen 1S--2P transition, with hydrogen being
a standard system for Lamb shift measurements and the
1S--2P transition as a very good 
realization of the two-level approximation.
If we assume a tightly focused laser beam (limit on the beam waist is
of the order of the laser wavelength), then a calculation shows
that the required Lyman-$\alpha$ power of $340\, \mu$W for
an $h$-parameter of $1000$ is less than $10^5$ times larger 
than the current maximum power of 20 nW~\cite{EiWaHa2001}.
Considerable progress (roughly a factor of $10^3$ in power)
might be achieved in the near future due to enhancement resonators
that are resonant to the frequencies of 
all three incoming laser beams contributing to the four-wave
mixing process which generates the coherent
Lyman-$\alpha$ light in mercury vapor~\cite{EiWaHa2001,WaPa2003}. 

In summary, we find that our calculated Lamb shift
of laser-dressed atomic states is nontrivially different from 
that via conventional approximate treatments
where the perturbative quantum electrodynamic interaction 
is evaluated prior to the exact quantum optical coupling 
with the laser field.
The corrections  are present even though the highly occupied 
laser mode, to a very good approximation (i.e., ignoring light-by-light
scattering) does not interact with the other vacuum modes 
which are responsible for the Lamb shift.
The radiative corrections amount 
to a change of the detuning corresponding to the Lamb shift
of the ``bare'' transition, and a radiative modification of the Rabi frequency.
The feasibility of a measurement on the hydrogenic 
1S--2P transition is discussed. Other interesting questions 
are related to classical analogues of two-level quantum systems
where the radiative corrections discussed necessarily assume
a fundamentally different form~\cite{BeSpAl1992}; these might be the
subject of forthcoming investigations.

Financial support by the German Science Foundation (SFB 276) is 
gratefully acknowledged.


\begin{thebibliography}{23}
\bibitem[{Rem()}]{RemarkLambShift}
\bibinfo{howpublished}{The Lamb shift of the 2P state, being of the order of
  $-$13 MHz only, may be neglected in comparison to the Lamb shift of the 1S
  state, which is of the order of $+$8000 MHz.}
\end{thebibliography}
\end{document}